# EARLY ADOPTION OF GENERATIVE AI BY GLOBAL BUSINESS LEADERS: INSIGHTS FROM AN INSEAD ALUMNI SURVEY

**JASON DAVIS**

INSEAD

1 Ayer Rajah Avenue

Singapore 138676

+65 6799 5257

jason.davis@insead.edu

**JIAN BAI "JAMBER" LI**

National University of Singapore

15 Kent Ridge Drive

Singapore 119245

bizlijb@nus.edu.sg

Working Paper

April 6, 2024

We are grateful for support from the research team including especially research assistant Ethan Gail, as well as the INSEAD Alumni Relations team including Zeina Sleiman, William Walsh, and Austin Tomlinson for supporting this research. Research support was also provided by the INSEAD Harborne Blockchain Fund.

# Early Adoption of Generative AI by Global Business Leaders: Insights from an INSEAD Alumni Survey

**ABSTRACT**

How are new technologies like generative AI quickly adopted and used by executive and managerial leaders to create value in organizations? A survey of INSEAD's global alumni base revealed several intriguing insights into perceptions and engagements with generative AI across a broad spectrum of demographics, industries, and geographies. Notably, there's a prevailing optimism about the role of generative AI in enhancing productivity and innovation, as evidenced by the 90% of respondents being excited about its time-saving and efficiency benefits. Despite this enthusiasm, concerns are significant, particularly regarding misuse by individuals (82%) and issues related to surveillance and privacy (67%). The adoption of generative AI is widespread, with 52% of organizations already using it and a majority of individuals incorporating it into both personal and professional realms. Analysis revealed different attitudes about adoption and use across demographic variables. Younger respondents are significantly more excited about generative AI and more likely to be using it at work and in personal life than older participants. Those in Europe have a somewhat more distant view of generative AI than those in North America in Asia, in that they see the gains more likely to captured by organizations than individuals, and are less likely to be using it professional and personal contexts than those in North America and Asia. This may also be related to the fact that those in Europe are more likely to be working in Financial Services and less likely to be working in Information Technology industries than those in North America and Asia. Despite this, those in Europe are more likely to see AGI happening faster than those in North America, although this may reflect less interaction with generative AI in personal and professional contexts. These findings collectively underscore the complex and multifaceted perceptions of generative AI's role in society, pointing to both its promising potential and the challenges it presents.

How new technologies are adopted by individuals and organizations is a topic of great importance because the diffusion of new technologies determines how quickly they find their best use and improve lives (Tushman and Anderson 1986; Tripsas 2009; Naumovska, Gaba, and Greve 2021). The development of large language models (LLMs) and emergence of generative artificial intelligence (AI) services offers an important lens with which to examine new technology diffusion and use. Although LLMs have a long history, it was the release of OpenAI's chatbot and chatGPT3.5 model in late 2022 that led to mass adoption of these technologies and the emergence of many rival LLMs by bigtech companies like Google, Microsoft, and Facebook, as well as notable open source efforts as well.

The diffusion of generative AI technologies to individuals and organizations has been rapid. ChatGPT was, famously, the digital platform with the fastest adoption curve, reaching 100 million users in only a few months (OpenAi 2023). Business organizations have been particularly quick to adopt LLMs and related generative AI services, although other institutions and professions such as medicine (Lee, Goldberg, and Kohane 2023) and education (Mollick and Mollick 2023) have been at the forefront of use as well. For instance, a survey of 225 US executives indicated that 65% believe generative AI will have a high or extremely high impact on their organization in the next 3-5 years. And 60% say they are 1-2 years away from implementing their first generative AI solution. Executives are most optimistic about opportunities to increase productivity (72%), change the way people work (65%), and encourage innovation (66%) (KPMG 2023).

Generative AI technologies appear to be valuable, even if there are some notable concerns. For example, LLMs have been shown to improve productivity in a variety of work settings ranging from the automation of routine tasks (Agrawal, Gans, and Goldfarb 2023) to strategic decision making (Gaessler and Piezunka 2023; Lebovitz, Lifshitz-Assaf, and Levina 2022; Doshi 2024) to entrepreneurship (Otis et al. 2023) to creativity (Doshi and Hauser 2024; Girotra et al. 2023; Jia et al. 2023; Mukherjee and Chang 2023). There has been considerable effort dedicated to exploring how generative AI will impact jobs and work tasks in organizations (Dell'Acqua et al. 2023; Eloundou et al. 2023; Brynjolfsson, Li, and Raymond 2023; Felten, Raj, and Seamans 2023), with some finding positive impact on work productivity and new jobs (Chomsky, Roberts, and Watumull 2023; Otis et al. 2023) while other have found job loss (Bernd Carsten Stahl 2023; Hui 2023). Ethical concerns have also been raised about the utility and bias stemming from training sets (Hannigan, McCarthy, and Spicer 2023; Lebovitz, Levina, and Lifshitz-Assaf 2021; Chomsky, Roberts, and Watumull 2023), the impact on employee agency (Vanneste and Puranam 2024; Ali et al. ; Raisch and Krakowski 2021), and inequality in who will capture the value of generative AI (Berg, Raj, and Seamans 2023), whether it is the owners of capital vs employees, the organization itself versus individual members, or top executives versus those in lower ranks.

It is important to understand is how leaders of organizations – both top executives and mid-level managers – are adopting and using these technologies, including their attitudes towards the risk and benefits, their perceptions of who benefits, and what the future may be, because they will strongly shape the impact of generative AI. It may particularly useful to understand how organizational leaders approach new technologies like generative AI early in its adoption phase where the impact is ambiguous. Of particular importance is the organization's stance towards future technology evolution and the emergence of superhuman AI and artificial general intelligence (AGI), as corporate adoption may shape the future of humanity and social life (Boussioux et al. 2023; Beane 2019; Davis 2023).

The adoption and use of these technologies may differ substantially across organizations in differ industries and countries, and individual employees may have different perceptions depending on their role in the organization and position in the hierarchy. Generative AI is thought to be a general purpose technology (Bresnahan and Tratjenberg 1995; Goldfarb 2005) – that is, of wide use in many applications for many organizations and people. Yet these variations in adoption and use may be what determines the locus of value creation in these areas. A major difficulty is that that vast majority of public discussion and research about generative AI has focused on applications in the United States and the technology sector where they were developed, leaving a wide swath of application areas underexplored.

The purpose of this paper is to examine how generative AI technologies are being adopted and used by individual leaders and organizations in various industries and geographies. We leverage a survey of alumni from INSEAD, an international business school – INSEAD's alumni base is unique in its global distribution, with over 60000 executives, managers, and founders of top organizations in over 180 countries worldwide. This global survey allows us to examine attitudes of business leaders towards generative AI, their concerns, and use, including their perceptions of who is capturing the value of generative AI, and what the future of generative AI may bring for organizations and individual employees.

## METHODS

The primary aim of this study is to explore the diffusion of emergent generative AI technology, particularly during its initial rollout phase. This includes examining its perceived impacts on society, the business sector, and individual careers, as well as its usage patterns. Of particular interest is how generative AI adoption and use differs across industries and regions, and how it is impacting executives, managers, and employees in organizations. The choice to study the adoption of generative AI among INSEAD alumni stems from several advantages. Established in 1957, INSEAD is a global business school with a presence in Europe, Asia, the Middle East, and North America, boasting a worldwide alumni network. INSEAD alumni are among the world's most influential business leaders, holding executive positions in many leading global corporations listed in various indexes such as the FTSE 100, MSCI world index, and S&P 500. INSEAD is also renowned for entrepreneurship, with alumni having founded over 1,400 companies and raised $35 billion (INSEAD 2023), notably achieving a significant footprint in Europe as the academic institution with the highest number of European unicorns founded by its alumni.

The survey was sent to all INSEAD alumni, a diverse group of 62870 individuals residing in 179 countries in July 2023. Alumni surveys serve as a valuable tool for collecting data from a broad population across various industries, noted for higher response rates and trust levels compared to other populations (Eesley 2011; Lazear 2004; Burt 2001; Eesley, Li, and Yang 2012). The survey was available for responses during July and August 2023, focusing on gathering early insights into the adoption, use cases, and perceived impacts of generative AI. Out of 40,000 alumni contacted, we received 1,207 usable responses, indicating an 8% response rate. However, the response count was lower for certain questions due to unusable or blank submissions, leading to reduced sample sizes for gender, location, and age (1,046), industry (1,048), and organization size (1,037).

The respondent demographic is diverse: 61% reside in Europe, 20% in Asia, and 11% in North America. A significant portion, 79%, are aged between 31 and 60 years. While responses span across various industries, information technology and financial services are the most represented, accounting for 43% of the total. Additionally, respondents are distributed across different organizational sizes, with 43% working in organizations with more than 1,000 employees. Gender distribution among respondents is predominantly male (75%) with females constituting 23%. The survey also highlights a skew towards senior roles within organizations, with 68% of participants identifying as executives, top managers, or managers, and 41% as owners or board members.

Demographics and Summary Statistics

The population of INSEAD alumni with active email addresses includes 62870 – we sent our survey to them all. This population is known for being highly geographically distributed, representing over 179 countries of residence, wither over 54% in Europe, 18% in Asia Pacific, and 8% in North America. Approximately 65% are between 31 and 60. 79% are Male. Approximately half of alumni are from INSEAD's MBA degree program, while the remaining are mostly from INSEAD's executive education programs (INSEAD 2024). As will be described below, the responding sample is roughly similar on these demographic variables.

Below are summary statistics about the sample of survey respondents. I report detailed demographic findings about gender, organization size, industry, age, organization position, and region of survey respondents before moving to the core results.

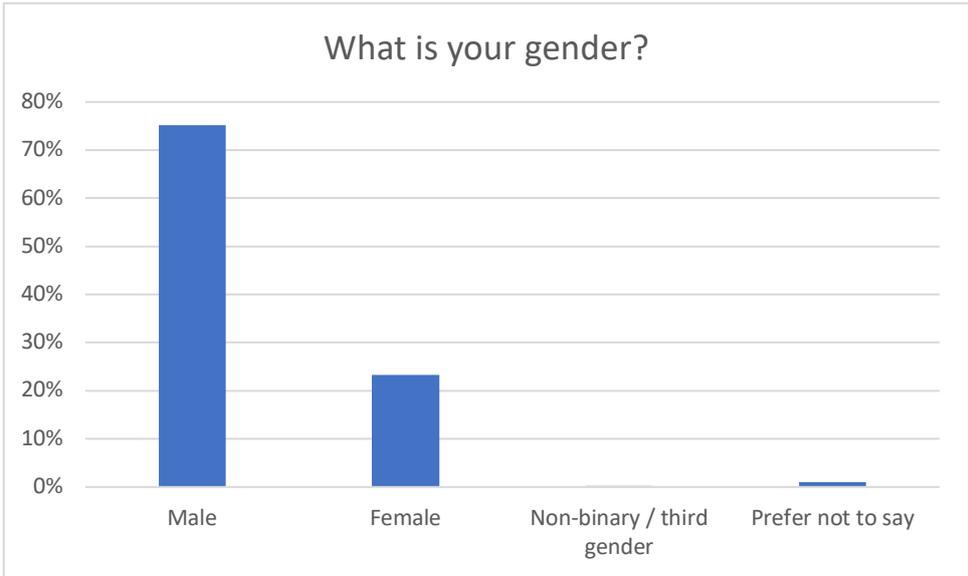

| Gender | What is your gender? |
|---|---|
| Male | 75% |
| Female | 23% |
| Non-binary / third gender | 0% |
| Prefer not to say | 1% |

The survey included a question regarding the gender identity of participants to better understand the demographic composition of respondents. The majority, 75%, identified as male, while 23% identified as female. Notably, none of the respondents selected the option for non-binary or third gender, and a small fraction of 1% chose to prefer not to say. This demographic distribution suggests a predominance of male participants in the survey.

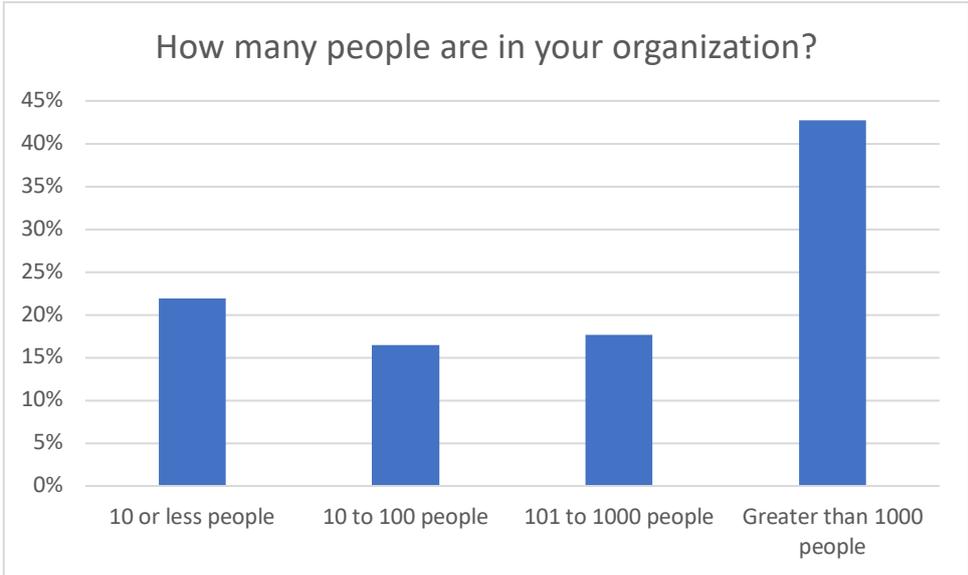

| Organization Size | How many people are in your organization? |
|---|---|
| 10 or less people | 22% |
| 10 to 100 people | 16% |
| 101 to 1000 people | 18% |
| Greater than 1000 people | 43% |

The survey inquired about the size of the organizations that participants belong to, revealing a diverse array of organizational scales. The largest proportion, 43%, reported being part of organizations with more than 1000 people, indicating that a significant number of respondents are engaged in large-scale enterprise environments. On the other end of the spectrum, 22% belong to small entities comprising 10 or fewer individuals, highlighting the participation of startups and small businesses. Those in organizations with 10 to 100 people and 101 to 1000 people represent 16% and 18% of respondents, respectively, showcasing involvement from mid-sized organizations. This distribution underscores the wide-ranging applicability and interest in the survey topic across different organizational sizes, from small teams to large corporations.

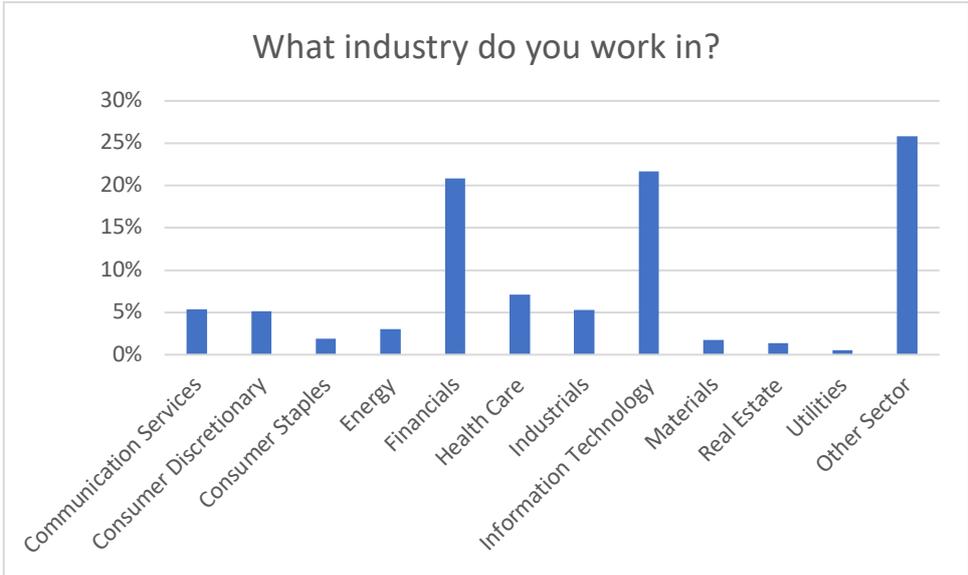

| Industry | What industry do you work in? |
|---|---|
| Communication Services | 5% |
| Consumer Discretionary | 5% |
| Consumer Staples | 2% |
| Energy | 3% |
| Financials | 21% |
| Health Care | 7% |
| Industrials | 5% |
| Information Technology | 22% |
| Materials | 2% |
| Real Estate | 1% |
| Utilities | 1% |
| Other Sector | 26% |

The survey asked participants about the industry sector of their employment to identify where interest in generative AI is most concentrated. The most represented sector was 'Other Sector' at 26%, suggesting a wide interest across various fields not listed explicitly or indicating niche industries. This was closely followed by Information Technology, with 22% of respondents, highlighting the strong engagement of this sector with AI technologies. Financials also showed significant representation at 21%, underscoring the sector's interest in leveraging AI for improved decision-making and efficiency. Health Care and Communication Services each garnered 7% and 5% respectively, reflecting a diverse interest in AI across sectors. Notably, sectors such as Consumer Staples, Materials, Real Estate, and Utilities had minimal representation, each accounting for 2% or less. This distribution illustrates the broad interest in generative AI across a range of industries, with a particular emphasis on technology and finance sectors.

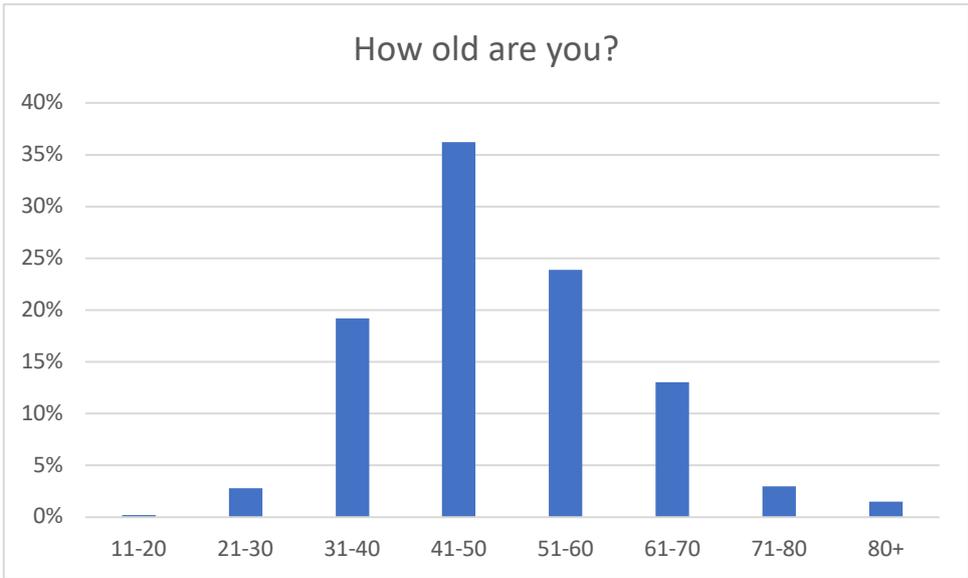

| Age | How old are you? |
|---|---|
| 11-20 | 0% |
| 21-30 | 3% |
| 31-40 | 19% |
| 41-50 | 36% |
| 51-60 | 24% |
| 61-70 | 13% |
| 71-80 | 3% |
| 80+ | 2% |

The survey collected data on the ages of participants to better understand the demographics of those interested in or affected by generative AI. The results show a broad age range among respondents, with the largest group, 36%, falling within the 41-50 age bracket, suggesting a mature audience with potentially significant professional experience is engaging with AI discussions. The next largest age groups are 51-60 and 31-40 years old, with 24% and 19% of respondents respectively, indicating a strong interest in AI technologies among middle-aged professionals. Those in the 61-70 age range represent 13%, while the younger 21-30 and older 71-80 age brackets each comprise 3% of participants, with those over 80 years old making up 2%. Notably, there were no participants in the 11-20 age group, highlighting a gap in engagement among the youngest demographic. This age distribution underscores the appeal of generative AI topics primarily among those in the mid to late stages of their careers.

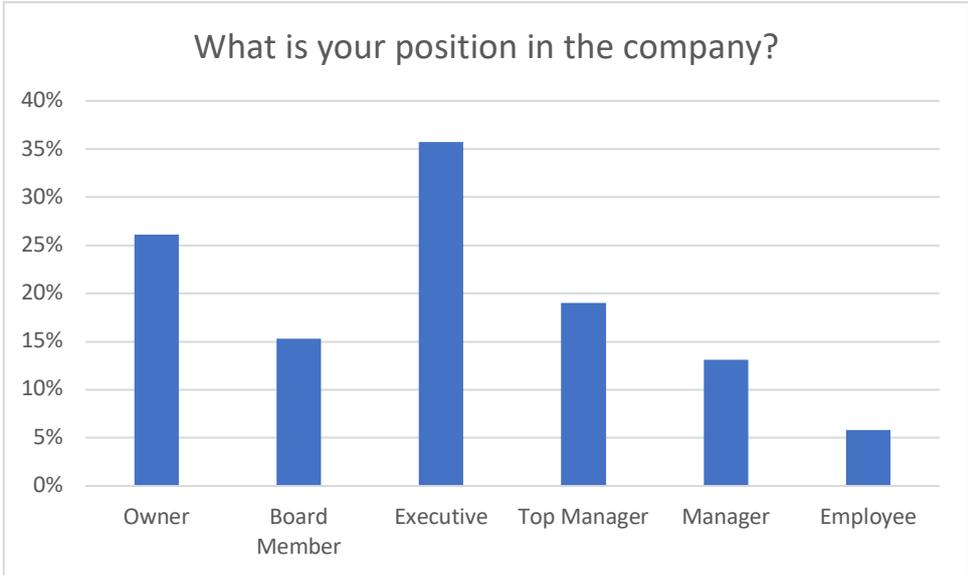

| Position / Role | What is your position in the company? |
|---|---|
| Owner | 26% |
| Board Member | 15% |
| Executive | 36% |
| Top Manager | 19% |
| Manager | 13% |
| Employee | 6% |

The survey queried respondents about their roles within their companies to ascertain the levels of interest in generative AI across various organizational positions. The data reveals a significant representation from those in leadership positions, with Executives making up the largest group at 36%, highlighting a keen interest in AI technologies among top-level decision-makers. Owners accounted for 26% of responses, indicating a strong engagement from those with a vested interest in the strategic direction of their businesses. Top Managers and Board Members also showed notable interest, with 19% and 15% respectively, suggesting that the implications of generative AI are being considered at all levels of leadership. Managers and Employees, representing 13% and 6% of the survey population respectively, demonstrate that while interest in AI extends across the organizational hierarchy, it is most pronounced among those with the highest levels of responsibility. It should be noted that respondents can select multiple positions. This distribution points to a significant focus on generative AI among those in positions to influence organizational strategy and innovation.

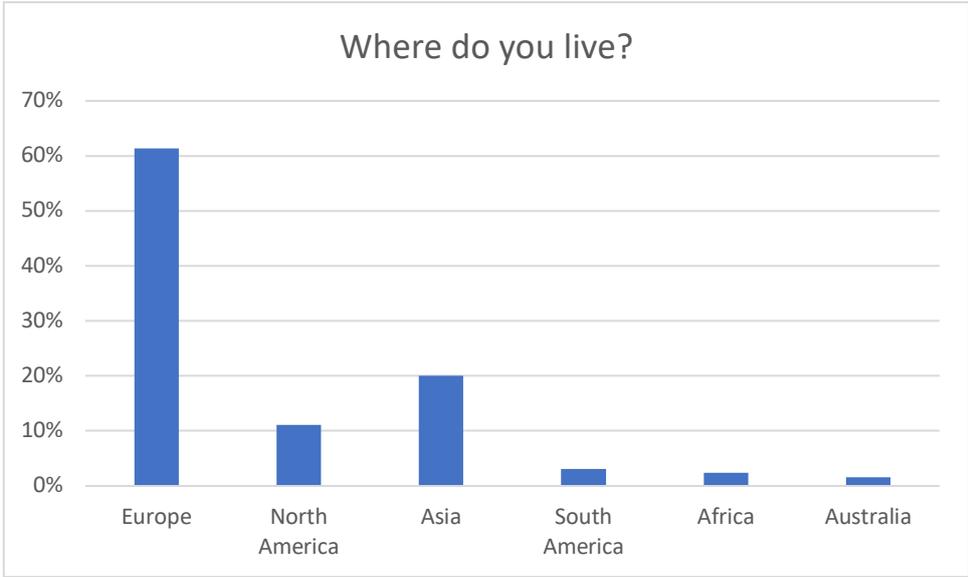

| Location | Where do you live? |
|---|---|
| Europe | 61% |
| North America | 11% |
| Asia | 20% |
| South America | 3% |
| Africa | 2% |
| Australia | 2% |

The survey investigated the geographical distribution of respondents to identify global perspectives on generative AI. A substantial majority, 61%, are based in Europe, indicating a strong interest and engagement with AI technologies within this region. Asia accounts for 20% of the respondents, showcasing significant involvement from this diverse and technologically advancing area. North America, while having a notable AI development scene, represents 11% of the survey population, suggesting the survey might have had more limited reach or different engagement levels in this region. Contributions from South America, Africa, and Australia are relatively minimal, each constituting 2-3% of responses, highlighting an opportunity for increased global participation and awareness in discussions surrounding AI. This distribution underscores the predominance of European interest in the survey, while also reflecting a varied international perspective on generative AI.

**RESULTS**

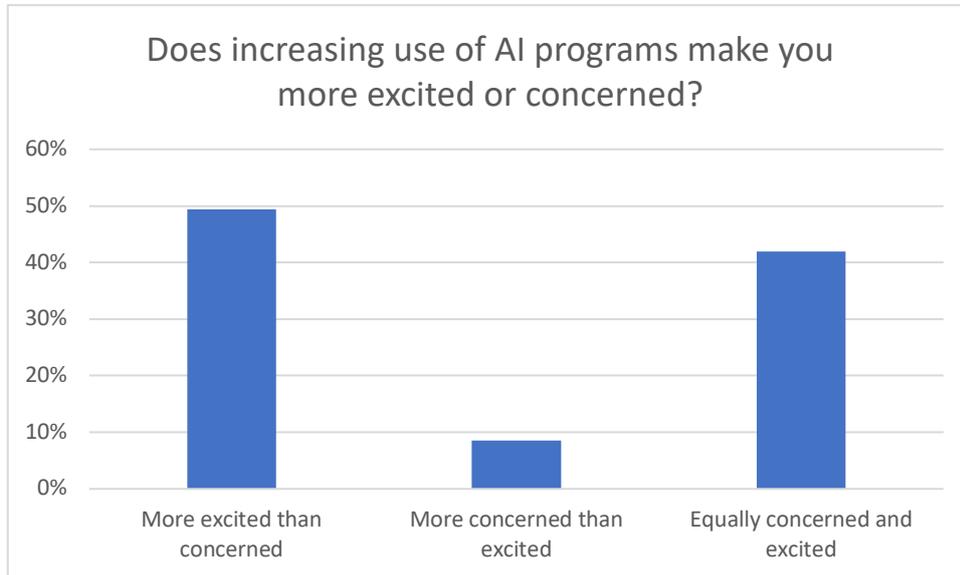

| Excitement vs Concern | Does increasing use of AI programs make you more excited or concerned? |
|---|---|
| More excited than concerned | 49% |
| More concerned than excited | 9% |
| Equally concerned and excited | 42% |

This survey question aimed to understand how people feel about the increasing use of AI programs. Before focusing on generative AI, it was crucial to assess general attitudes. The question asked was, "Does increasing use of AI programs make you more excited or concerned?" The majority, 49% of respondents, reported being "more excited than concerned" about the rise of AI programs. A similar portion, 42%, expressed being "equally concerned and excited," indicating a balanced view. Only 9% felt "more concerned than excited," showing that concerns are relatively minor compared to the excitement or balanced views. Overall, the results suggest a predominantly positive outlook towards the increasing use of AI programs, with most respondents leaning towards excitement or holding a balanced perspective.

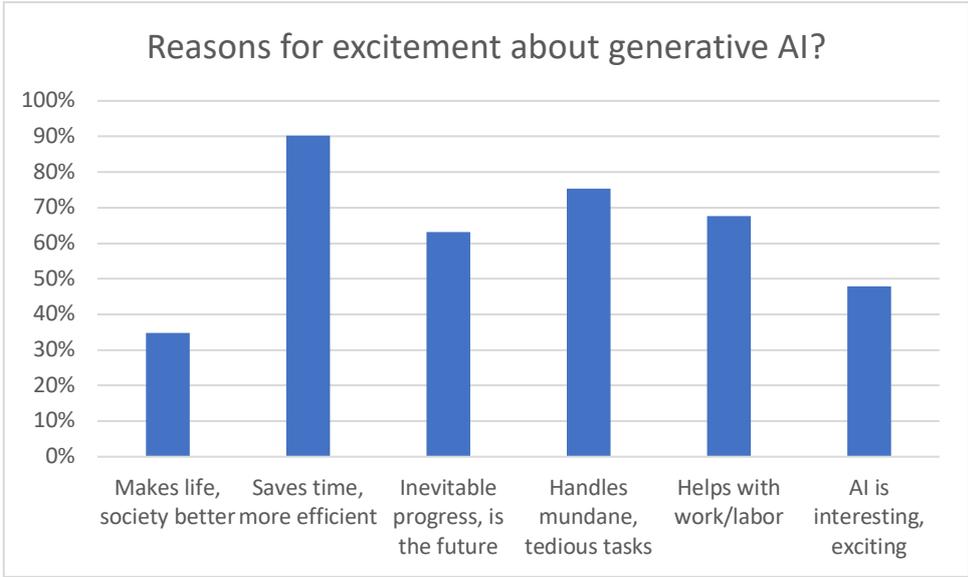

| Excitement | Reasons for excitement about generative AI? |
|---|---|
| Makes life, society better | 35% |
| Saves time, more efficient | 90% |
| Inevitable progress, is the future | 63% |
| Handles mundane, tedious tasks | 75% |
| Helps with work/labor | 68% |
| AI is interesting, exciting | 48% |

The survey explored specific reasons behind the excitement for generative AI among respondents. The leading reason, with 90% agreement, was that generative AI saves time and increases efficiency, highlighting the practical benefits of these technologies. Close behind, 75% of participants appreciated generative AI for handling mundane and tedious tasks, with 68% acknowledging its assistance with work or labor, indicating a strong valuation of AI's capacity to enhance productivity and workplace dynamics. Furthermore, 63% of respondents viewed generative AI as inevitable progress and the future, reflecting a recognition of its transformative potential. AI's inherent interest and excitement were noted by 48%, while 35% believed it would make life and society better, suggesting optimism about its broader impacts. These results underscore a widespread enthusiasm for generative AI, driven by its perceived utility, efficiency, and revolutionary role in future progress.

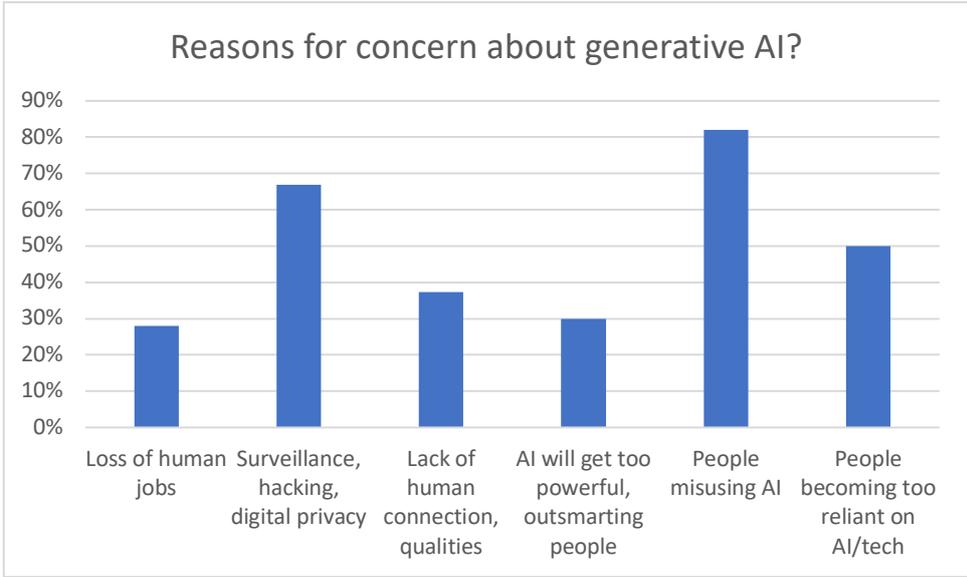

| Concerns | Reasons for concern about generative AI? |
|---|---|
| Loss of human jobs | 28% |
| Surveillance, hacking, digital privacy | 67% |
| Lack of human connection, qualities | 37% |
| AI will get too powerful, outsmarting people | 30% |
| People misusing AI | 82% |
| People becoming too reliant on AI/tech | 50% |

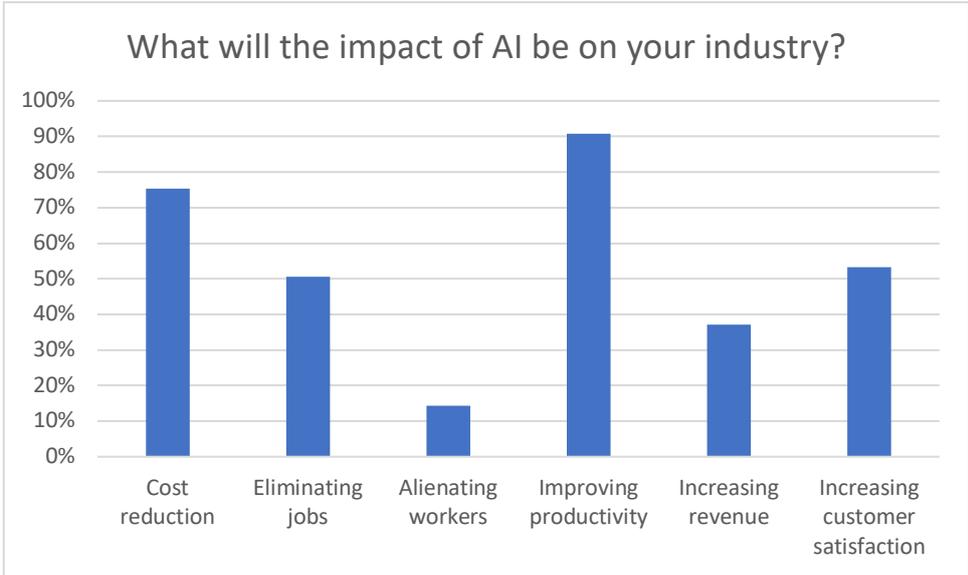

| Industry Impacts of Generative AI | What will the impact of AI be on your industry? |
|---|---|
| Cost reduction | 75% |
| Eliminating jobs | 51% |
| Alienating workers | 14% |
| Improving productivity | 91% |
| Increasing revenue | 37% |
| Increasing customer satisfaction | 53% |

This survey question delved into the concerns surrounding generative AI, revealing a spectrum of apprehensions. The most prominent concern, shared by 82% of respondents, was the potential for people to misuse AI, indicating widespread anxiety about the ethical and safety implications of these technologies. Surveillance, hacking, and issues around digital privacy were also major concerns for 67% of participants, highlighting fears about the erosion of privacy and security in the age of AI. Half of the respondents worried about people becoming too reliant on AI and technology, reflecting concerns about dependency and the loss of autonomy. Concerns about AI leading to a loss of human jobs, outsmarting humans, and diminishing human connection were less prevalent, cited by 28%, 30%, and 37% of participants, respectively. These findings suggest that while there are significant concerns about the misuse and ethical implications of generative AI, fears about job loss and AI's autonomy are relatively lower on the list of worries.

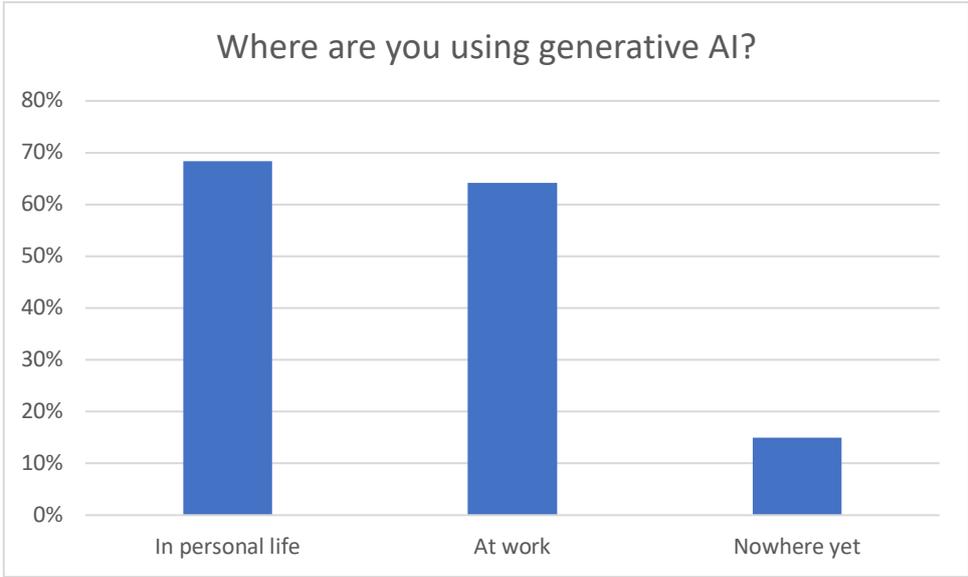

| Individual Generative AI Usage | Where are you using generative AI? |
|---|---|
| In personal life | 68% |
| At work | 64% |
| Nowhere yet | 15% |

The survey also investigated where individuals are currently utilizing generative AI. A significant majority, 68%, reported using generative AI in their personal lives, indicating its widespread acceptance and integration into daily routines and activities. Close behind, 64% of respondents are applying generative AI at work, suggesting its growing influence in professional environments and its potential to enhance productivity and creativity in the workplace. However, a notable minority of 15% have not yet engaged with generative AI technologies, pointing to either a lack of access, interest, or awareness of how these tools can be applied in their lives. These results highlight the rapid adoption of generative AI across both personal and professional spheres, yet also acknowledge a segment of the population that remains untouched by this technological wave.

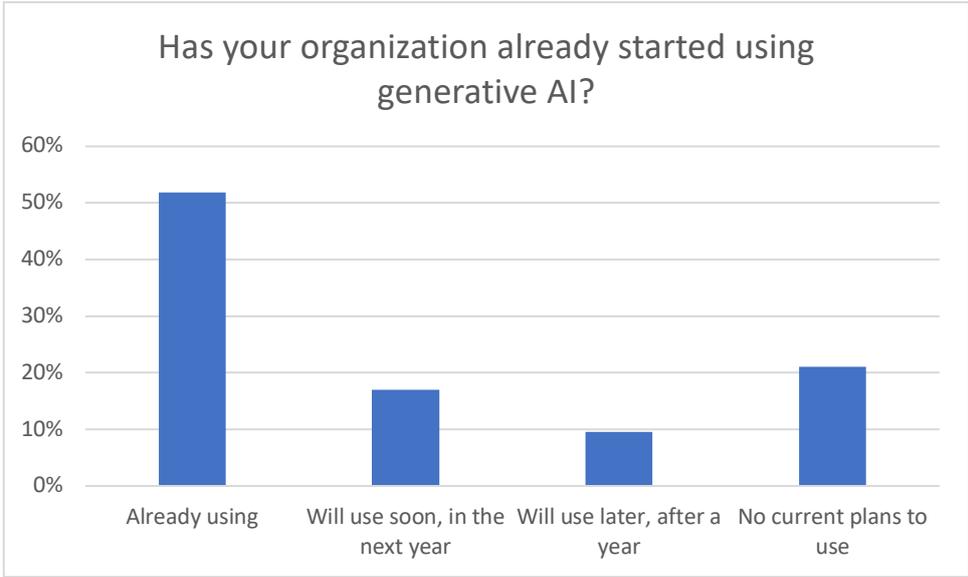

| Organization Generative AI Usage | Has your organization already started using generative AI? |
|---|---|
| Already using | 52% |
| Will use soon, in the next year | 17% |
| Will use later, after a year | 10% |
| No current plans to use | 21% |

The survey sought insights into the adoption of generative AI within organizations, revealing that a majority, 52%, are already utilizing these technologies. This indicates a significant uptake and integration of generative AI in business operations, underscoring its perceived value and utility. Additionally, 17% of organizations plan to start using generative AI within the next year, while a further 10% anticipate adoption after a year, suggesting a growing interest and planned investment in these technologies. However, 21% of organizations reported no current plans to engage with generative AI, reflecting either skepticism, a wait-and-see approach, or satisfaction with current tools and processes. Overall, these findings demonstrate a strong and increasing engagement with generative AI in the organizational context, albeit with a notable minority yet to be convinced of its benefits.

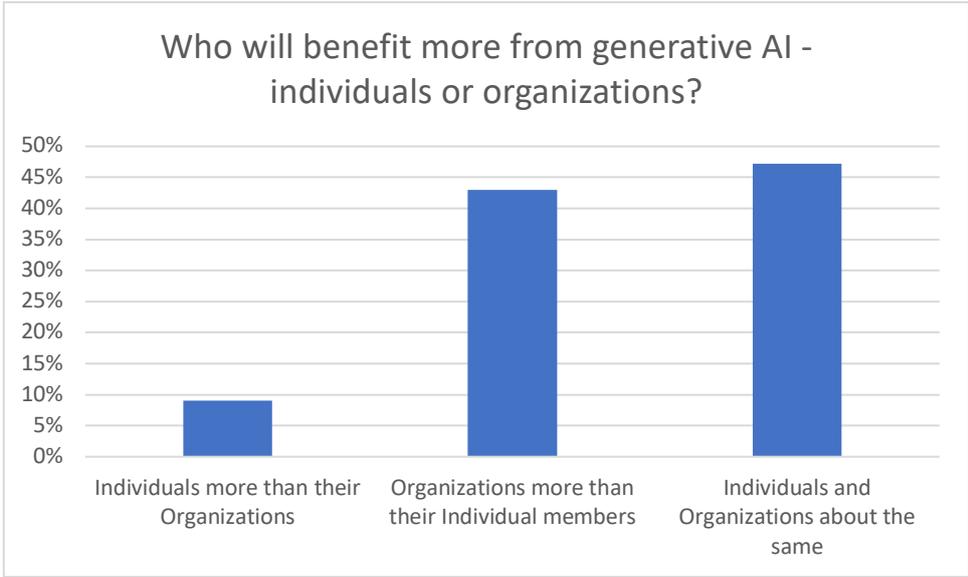

| **Individuals vs Organizations** | **Who will benefit more from generative AI - individuals or organizations?** |
|---|---|
| Individuals more than their Organizations | 9% |
| Organizations more than their Individual members | 43% |
| Individuals and Organizations about the same | 47% |

This survey question addressed perspectives on who stands to gain more from generative AI, whether individuals or organizations. A plurality of respondents, 47%, believe that both individuals and organizations will benefit approximately equally from generative AI, highlighting the perceived widespread advantages of these technologies across different sectors of society. However, 43% feel that organizations will benefit more than their individual members, suggesting that the scalability and efficiency gains offered by generative AI are seen as particularly advantageous for businesses and other formal structures. Only a small fraction, 9%, view individuals as benefiting more than organizations, indicating a perception that the personal utility of generative AI, while significant, may not match the transformative impact it can have on organizational performance and productivity. These findings reflect a consensus that generative AI holds substantial promise for both individuals and organizations, with a slight inclination towards greater benefits for organizational entities.

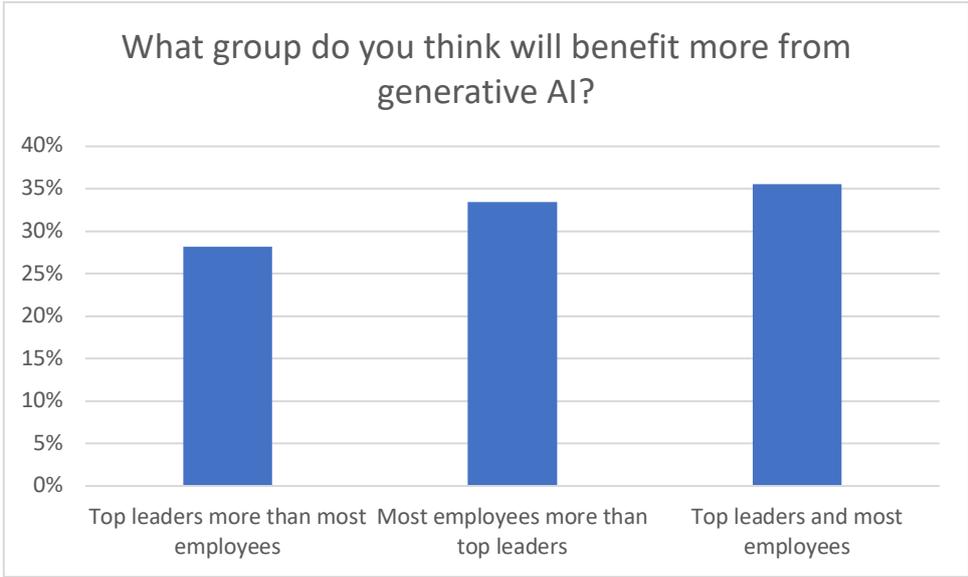

| Top Leaders vs Most Employees | What group do you think will benefit more from generative AI? |
|---|---|
| Top leaders more than most employees | 28% |
| Most employees more than top leaders | 33% |
| Top leaders and most employees | 36% |

The survey explored perceptions on which group within organizations—top leaders or most employees—would benefit more from the adoption of generative AI. A slight majority of 36% of respondents believe that both top leaders and most employees will equally benefit from generative AI, suggesting a view that its advantages can permeate through all levels of an organization. Interestingly, 33% feel that most employees will derive more benefit than top leaders, potentially reflecting opinions that generative AI will democratize access to information, improve efficiency, and enhance task completion for a wider range of roles. Conversely, 28% opine that top leaders will benefit more, possibly due to the strategic advantage and decision-making support that generative AI offers. These results indicate a nuanced understanding of generative AI's impact, recognizing its potential to support various organizational roles differently yet beneficially.

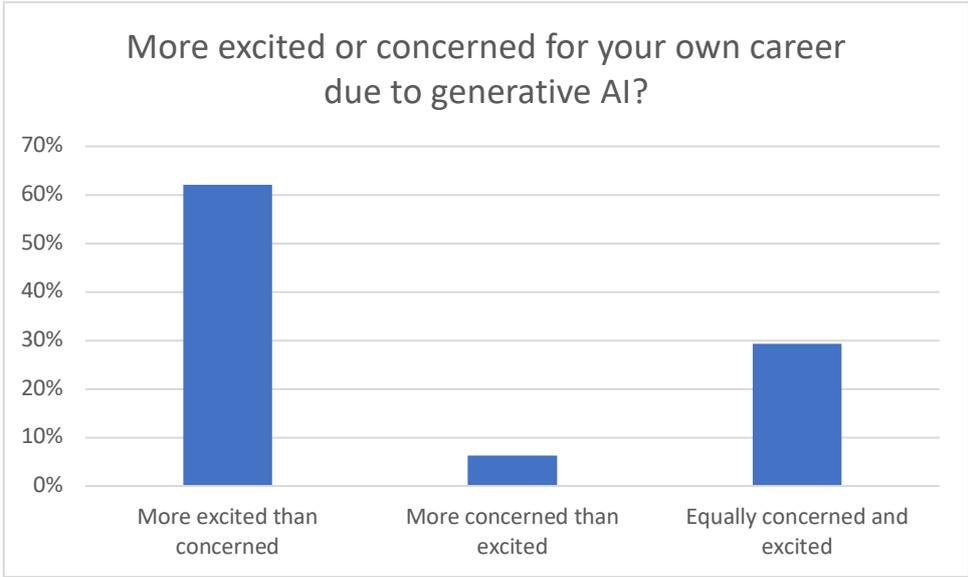

| Excited vs Concerned | More excited or concerned for your own career due to generative AI? |
|---|---|
| More excited than concerned | 62% |
| More concerned than excited | 6% |
| Equally concerned and excited | 29% |

The survey probed into personal sentiments regarding the impact of generative AI on individuals' careers, revealing a clear trend towards optimism. A significant 62% of respondents are more excited than concerned about the influence of generative AI on their career prospects, indicating a strong belief in the positive opportunities and advancements that AI technologies can bring to their professional development. Only a small fraction, 6%, expressed more concern than excitement, perhaps wary of the disruptions and challenges AI might pose. Meanwhile, 29% of participants hold a balanced view, being equally concerned and excited, which suggests an awareness of both the potential benefits and uncertainties associated with generative AI in the workplace. Overall, these findings underscore a predominantly positive outlook among professionals regarding the role of generative AI in shaping future career landscapes.

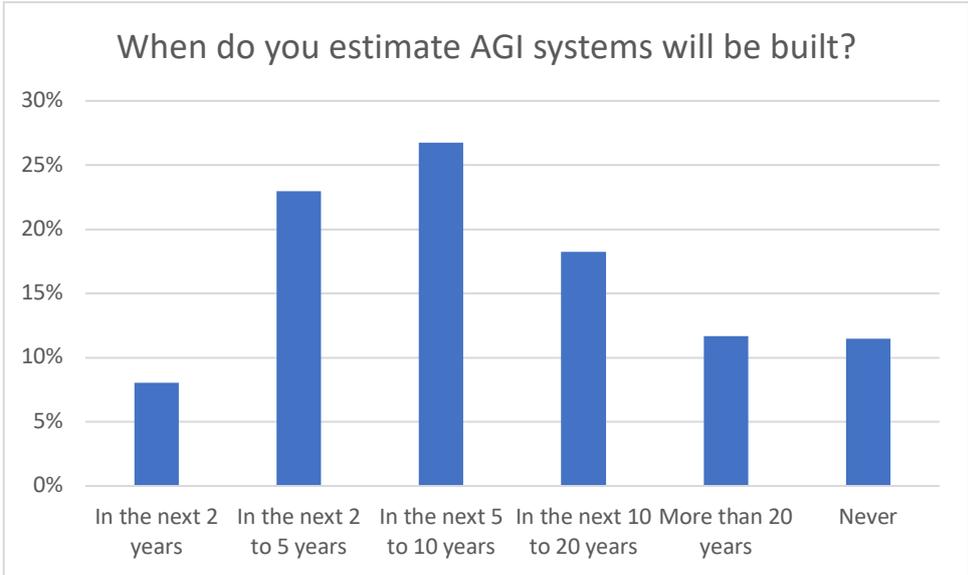

| Time to AGI | When do you estimate AGI systems will be built? |
|---|---|
| In the next 2 years | 8% |
| In the next 2 to 5 years | 23% |
| In the next 5 to 10 years | 27% |
| In the next 10 to 20 years | 18% |
| More than 20 years | 12% |
| Never | 11% |

The survey sought opinions on the anticipated timeline for the development of Artificial General Intelligence (AGI) systems. Responses indicate a spread of expectations, with the most common estimate, held by 27% of participants, being that AGI will emerge in the next 5 to 10 years. This is followed by 23% of respondents predicting the development of AGI systems within the next 2 to 5 years, and 18% foreseeing it happening in the next 10 to 20 years. A smaller group, 8%, are more optimistic, believing AGI could be realized in the next 2 years, whereas 12% think it will take more than 20 years. Interestingly, 11% of respondents are skeptical about the feasibility of AGI, believing it will never be achieved. These results reflect a wide range of perspectives on the pace of AI advancements, with a general consensus leaning towards significant progress within the next decade.

**Crosstabs and Demographic Analysis**

Next, we turn attention to examining any differences in responses across participants with different demographic characteristics, sometimes referred to as the "crosstabs" in the dataset. For these analyses, we use Chi-square analyses to examine statistical significance. We report these tests below.

Next, we examine the attitudes of alumni of different ages towards generative AI. Participants were asked, "Does increasing use of AI programs make you more excited or concerned?" and were given three options to express their level of excitement or concern. The data indicates a general trend of excitement or neutrality rather than concern about generative AI across all age groups. However, a particularly interesting outcome is that younger respondents show a higher level of excitement compared to older respondents. For instance, 65.5% of individuals aged 21-30 are "More excited than concerned," in contrast to just 38.2% of those aged 61-70 who feel the same.

| | | **How old are you?** | | | | | | | |
|---|---|---|---|---|---|---|---|---|---|
| | | 11-20 | 21-30 | 31-40 | 41-50 | 51-60 | 61-70 | 71-80 | 80+ |
| **Does increasing use of AI programs make you more excited or concerned?** | More excited than concerned | 33.3% | 65.5% | 61.2% | 47.8% | 49.8% | 38.2% | 38.7% | 18.8% |
| | More concerned than excited | 33.3% | 3.4% | 6.0% | 7.1% | 7.2% | 12.5% | 16.1% | 37.5% |
| | Equally concerned and excited | 0.0% | 31.0% | 32.8% | 44.9% | 43.0% | 49.3% | 41.9% | 43.8% |

A chi-square test of significance was conducted, yielding a Chi-square statistic of 51.627 with a p-value of approximately 0.00000325 and 14 degrees of freedom. The p-value is significantly below the commonly accepted alpha level of 0.05, leading us to reject the null hypothesis. This implies that there is a statistically significant association between the level of excitement or concern about AI programs and the age group of the respondents. In summary, the analysis highlights a clear relationship between age and attitudes towards AI, with younger generations feeling more optimistic about the rise of generative AI technologies.

We now turn our attention to the domains in which alumni of various ages are employing generative AI. When asked "In which domains are you using generative AI?", respondents could indicate using AI in their personal life, at work, or report not using it yet. The data reveals significant use of generative AI in personal lives amongst younger age groups, with 72.4% of respondents aged 21-30 and 78.1% of those aged 31-40 incorporating AI into this domain. However, for the oldest age bracket (80+), only 6.3% reported using generative AI personally. The use of AI at work is also notable, particularly amongst the 21-30 age group (86.2%), and drops significantly in older age groups, with only 12.5% of those over 80 applying AI in their work environment. The proportion of respondents who have not yet used generative AI increases with age, reaching 87.5% in the 80+ category.

| | | **How old are you?** | | | | | | | |
|---|---|---|---|---|---|---|---|---|---|
| | | 11-20 | 21-30 | 31-40 | 41-50 | 51-60 | 61-70 | 71-80 | 80+ |
| **In which domains are you using generative AI?** | In personal life | 0.0% | 72.4% | 78.1% | 71.2% | 71.3% | 51.5% | 54.8% | 6.3% |
| | At work | 33.3% | 86.2% | 72.1% | 68.3% | 66.5% | 44.1% | 38.7% | 12.5% |
| | Nowhere yet | 0.0% | 3.4% | 6.0% | 11.6% | 12.4% | 32.4% | 29.0% | 87.5% |

A comprehensive chi-square test was conducted, yielding a Chi-square statistic of 175.255 and an extremely small p-value of approximately $5.925 \times 10^{-30}$ with 14 degrees of freedom. The p-value is substantially lower than the conventional alpha level of 0.05. Consequently, we can confidently reject the null hypothesis, affirming that there is a statistically significant association between the usage domains of generative AI and the respondents' age group. This underscores a discernible generational divide in the adoption and integration of generative AI technologies in various aspects of life.

The subsequent analysis explores the engagement of alumni of varying ages with generative AI within their organizations. The survey question posed was "Has your organization already started using Generative AI?" with options to indicate current use, plans to use within the next year, plans to use after a year, or no plans to use. The results show a strong current use of generative AI among the younger demographic, with 58.6% of 21-30-year-olds and 59.7% of 31-40-year-olds reporting that their organizations are already using it. Notably, there is a gradual decline in current usage as age increases, with only 12.5% of those aged 80+ reporting current use. Future plans to implement generative AI within the next year are highest among the 11-20 age group at 66.7%, while a substantial portion of the oldest age group (62.5%) indicates no plans to use generative AI.

| | | **How old are you?** | | | | | | | |
|---|---|---|---|---|---|---|---|---|---|
| | | 11-20 | 21-30 | 31-40 | 41-50 | 51-60 | 61-70 | 71-80 | 80+ |
| **Has your organization already started using Generative AI?** | Already using | 0.0% | 58.6% | 59.7% | 49.6% | 56.6% | 46.3% | 32.3% | 12.5% |
| | Will use soon, in the next year | 66.7% | 10.3% | 15.9% | 16.9% | 17.1% | 17.6% | 25.8% | 12.5% |
| | Will use later, after a year | 0.0% | 10.3% | 5.5% | 10.8% | 8.8% | 15.4% | 3.2% | 6.3% |
| | No current plans to use | 0.0% | 20.7% | 18.9% | 22.4% | 16.7% | 19.9% | 35.5% | 62.5% |

Statistical analysis was conducted using the chi-square test, resulting in a Chi-square statistic of 55.417 and a p-value of approximately 0.00006145, with 21 degrees of freedom. Given that the p-value is well below the standard alpha level of 0.05, we can reject the null hypothesis. This suggests a statistically significant relationship between the respondents' age groups and their organizations' plans for adopting generative AI. These findings underscore the generational differences in the adoption of AI technology within professional environments.

In this analysis, we delve into the perceptions within various industries regarding who stands to benefit more from generative AI, be it individuals or organizations. Surveyed industry professionals were presented with the question, "Who will benefit more from generative AI - individuals or organizations?" The response options allowed them to indicate a greater benefit for individuals, organizations, or equal benefits for both. Notably, in sectors like Communication Services, Financials, and Materials, a significant portion of respondents feel that organizations will benefit more than individuals, with responses like 44.6%, 53.2%, and 50.0%, respectively. Contrastingly, in the Health Care and Utilities sectors, a majority believe that individuals and organizations will benefit equally, with 56.8% and 66.7% respectively supporting this view.

| | | What industry do you work in? | | | | | |
|---|---|---|---|---|---|---|---|
| | | Communications | Consumer Staples | Energy | Financials | Health Care |
| Who will benefit more from generative AI - individuals or organizations? | Individuals more than their Organizations | 8.9% | 0.0% | 15.0% | 15.2% | 8.3% | 12.2% |
| | Organizations more than their Individual members | 44.6% | 50.0% | 50.0% | 42.4% | 53.2% | 31.1% |
| | Individuals and Organizations about the same | 46.4% | 50.0% | 35.0% | 36.4% | 38.5% | 56.8% |

| | | What industry do you work in? | | | | | |
|---|---|---|---|---|---|---|---|
| | | Industrials | Info Tech | Materials | Real Estate | Utilities | Other Sector |
| Who will benefit more from generative AI - individuals or organizations? | Individuals more than their Organizations | 10.9% | 10.5% | 5.6% | 7.1% | 0.0% | 8.5% |
| | Organizations more than their Individual members | 34.5% | 41.2% | 50.0% | 28.6% | 33.3% | 39.7% |

|  |  | Individuals and Organizations about the same | 54.5% | 47.8% | 38.9% | 64.3% | 66.7% | 50.7% |

Statistical analysis was performed using the chi-square test, resulting in a Chi-square statistic of 30.288 with a p-value of approximately 0.1116, across 22 degrees of freedom. With the p-value exceeding the alpha level of 0.05, we do not reject the null hypothesis. This indicates that, according to the data gathered, there is no statistically significant difference between industries in the perception of who will benefit more from generative AI—individuals or organizations. This suggests a generally uniform expectation across sectors regarding the impact of generative AI.

In the current analysis, we explore the expectations across different industries regarding the timeline for the construction of Artificial General Intelligence (AGI) systems. Industry professionals responded to the question, "When do you estimate AGI systems will be built?" with options ranging from within the next 2 years to never. The data shows a diversity of expectations, with the Energy sector being the most optimistic about short-term development (next 2 years) at 9.1%, while the Consumer Staples sector shows the highest skepticism with 25% believing AGI systems will never be built. Interestingly, a significant portion of the Utilities sector is in agreement about a longer timeline, with 66.7% estimating the development in the next 10 to 20 years, contrasting sharply with other sectors.

|  |  | What industry do you work in? | | | | | |
|---|---|---|---|---|---|---|---|
|  |  | Communications | Consumer | Staples | Energy | Financials | Health Care |
| **When do you estimate AGI systems will be built?** | In the next 2 years | 5.4% | 3.7% | 0.0% | 9.1% | 9.6% | 8.1% |
|  | In the next 2 to 5 years | 37.5% | 31.5% | 15.0% | 21.2% | 22.9% | 27.0% |
|  | In the next 5 to 10 years | 23.2% | 27.8% | 25.0% | 36.4% | 28.9% | 23.0% |
|  | In the next 10 to 20 years | 14.3% | 13.0% | 20.0% | 18.2% | 16.5% | 18.9% |
|  | More than 20 years | 7.1% | 7.4% | 10.0% | 12.1% | 10.1% | 13.5% |
|  | Never | 8.9% | 14.8% | 25.0% | 0.0% | 11.5% | 9.5% |

|  |  | What industry do you work in? | | | | | |
|---|---|---|---|---|---|---|---|
|  |  | Industrials | Info Tech | Materials | Real Estate | Utilities | Other Sector |
| **When do you estimate AGI systems will be built?** | In the next 2 years | 7.3% | 7.9% | 11.1% | 0.0% | 0.0% | 9.2% |
|  | In the next 2 to 5 years | 18.2% | 16.2% | 11.1% | 35.7% | 33.3% | 25.0% |

| | | | | | | |
|---|---|---|---|---|---|---|
| In the next 5 to 10 years | 29.1% | 28.1% | 27.8% | 14.3% | 0.0% | 25.7% |
| In the next 10 to 20 years | 20.0% | 24.1% | 27.8% | 21.4% | 66.7% | 13.2% |
| More than 20 years | 9.1% | 13.2% | 16.7% | 14.3% | 0.0% | 13.2% |
| Never | 16.4% | 10.1% | 5.6% | 14.3% | 0.0% | 12.9% |

The statistical evaluation, employing the chi-square test, resulted in a Chi-square statistic of 62.690 and a p-value of approximately 0.2223, with 55 degrees of freedom. The p-value surpasses the threshold alpha level of 0.05, leading us to fail to reject the null hypothesis. Therefore, the data suggests there is no statistically significant correlation between the industries in which respondents work and their predictions for when AGI systems will be realized. This implies that industry affiliation does not significantly influence perceptions about the timeline for AGI development.

We now investigate industry perspectives on which group within organizations is perceived to benefit more from generative AI. Respondents from various industries were asked, "What group do you think will benefit more from generative AI?" They could choose whether top leaders or most employees would benefit more, or if the benefits would be distributed equally. The results show varying opinions across industries, with the Financials sector tending to believe that top leaders will benefit more (37.2%), while the Real Estate and Utilities sectors are more inclined to think that most employees will gain more (55.6% and 33.3%, respectively). Notably, in the Utilities sector, a large majority (66.7%) view that both top leaders and most employees will benefit equally.

| | | What industry do you work in? | | | | | |
|---|---|---|---|---|---|---|---|
| | | Communications | Consumer | Staples | Energy | Financials | Health Care |
| **What group do you think will benefit more from generative AI?** | Top leaders more than most employees | 23.2% | 27.8% | 35.0% | 30.3% | 37.2% | 20.3% |
| | Most employees more than top leaders | 41.1% | 33.3% | 25.0% | 18.2% | 28.4% | 44.6% |
| | Top leaders and most employees | 32.1% | 33.3% | 35.0% | 51.5% | 31.7% | 35.1% |

| | | What industry do you work in? | | | | | |
|---|---|---|---|---|---|---|---|
| | | Industrials | Info Tech | Materials | Real Estate | Utilities | Other Sector |
| **What group do you think will benefit** | Top leaders more than most employees | 32.7% | 26.8% | 16.7% | 21.4% | 0.0% | 25.7% |
| more from generative AI? | Most employees more than top leaders | 30.9% | 32.0% | 55.6% | 57.1% | 33.3% | 34.2% |
| | Top leaders and most employees | 34.5% | 39.5% | 22.2% | 21.4% | 66.7% | 35.7% |

The chi-square statistical test reveals a Chi-square statistic of 33.305 with a p-value of approximately 0.0577 and 22 degrees of freedom. Given that the p-value is just above the conventional alpha level of 0.05, the null hypothesis is not rejected. This indicates that there is no statistically significant difference across industries regarding who is believed to benefit more from generative AI. However, the proximity of the p-value to the significance threshold suggests that while industry may not play a major role, the question is finely balanced and other factors not captured in this analysis may influence perceptions.

In this segment, we assess the sentiments regarding the increasing use of AI programs among individuals from various geographic regions. The survey participants were asked, "Does increasing use of AI programs make you more excited or concerned?" Across Europe, North America, Asia, South America, and Africa, a substantial proportion of respondents express more excitement than concern, with figures like 48.1% for Europe and peaking at 53.1% for South America. Interestingly, Australia shows a slightly lower excitement level at 41.2%. Those who are equally concerned and excited about AI form a significant group as well, with percentages hovering around 43% for Europe and even higher at 47.1% for Australia.

| | | Where do you live? | | | | | |
|---|---|---|---|---|---|---|---|
| | | Europe | North America | Asia | South America | Africa | Australia |
| Does increasing use of AI programs make you more excited or concerned? | More excited than concerned | 48.1% | 52.6% | 50.5% | 53.1% | 52.0% | 41.2% |
| | More concerned than excited | 8.5% | 11.2% | 8.1% | 3.1% | 4.0% | 11.8% |
| | Equally concerned and excited | 43.0% | 36.2% | 41.4% | 43.8% | 44.0% | 47.1% |

The chi-square test applied to these observations yields a statistic of 5.027 with a p-value of approximately 0.8894 and 10 degrees of freedom. The p-value is significantly higher than the standard alpha level of 0.05, leading us to fail to reject the null hypothesis. This suggests that, within this data set, there is no statistically significant difference in the levels of excitement or concern about AI programs based on the geographic location of the respondents. The general consensus appears to lean towards excitement or a balanced view about AI across different continents.

The analysis now shifts to the application of generative AI in different domains based on the geographic location of respondents. They were queried, "In which domains are you using generative AI?" with options to indicate usage in personal life, at work, or not at all yet. The majority of participants from all regions report using generative AI both in personal life and at work, with Asia leading in personal use at 79.0% and Africa matching this lead at work with 76.0%. The proportion of respondents who have not yet engaged with generative AI varies, with Asia having the lowest at 9.5% and South America the highest at 18.8%.

| | | **Where do you live?** | | | | | |
|---|---|---|---|---|---|---|---|
| | | Europe | North America | Asia | South America | Africa | Australia |
| **In which domains are you using generative AI?** | In personal life | 66.1% | 64.7% | 79.0% | 59.4% | 76.0% | 64.7% |
| | At work | 63.0% | 67.2% | 65.7% | 59.4% | 76.0% | 64.7% |
| | Nowhere yet | 16.4% | 16.4% | 9.5% | 18.8% | 8.0% | 11.8% |

The chi-square test used to examine the data produced a Chi-square statistic of 10.686 and a p-value of approximately 0.3825, with 10 degrees of freedom. As the p-value exceeds the typical alpha level of 0.05, we do not reject the null hypothesis. This suggests there is no statistically significant difference in the use of generative AI across different geographic locations as per the data collected. It appears that irrespective of where individuals reside, the uptake of generative AI in both personal and professional spheres is widespread, with only a small portion of the global respondents yet to begin using these technologies.

The focus of this analysis is on perceptions regarding who is likely to benefit more from generative AI across different global regions. Survey participants were questioned, "Who will benefit more from generative AI - individuals or organizations?" The responses show that in regions such as Europe, North America, and Asia, a significant portion of respondents believe that both individuals and organizations will benefit about the same, with Europe at 49.2% and Asia at 41.9%. In South America, this belief is even more pronounced, with 59.4% holding this view. In contrast, a majority in Australia (58.8%) feel that organizations will benefit more than individuals.

| | | **Where do you live?** | | | | | |
|---|---|---|---|---|---|---|---|
| | | Europe | North America | Asia | South America | Africa | Australia |
| **Who will benefit more from generative AI - individuals or organizations?** | Individuals more than their Organizations | 7.1% | 12.1% | 14.3% | 0.0% | 16.0% | 5.9% |
| | Organizations more than their Individual members | 43.0% | 42.2% | 43.3% | 40.6% | 40.0% | 58.8% |
| | Individuals and Organizations about the same | 49.2% | 44.0% | 41.9% | 59.4% | 44.0% | 35.3% |

Employing the chi-square test for analysis, we observe a Chi-square statistic of 19.225 with a p-value of approximately 0.0375 and 10 degrees of freedom. The p-value falls below the conventional alpha level of 0.05, enabling us to reject the null hypothesis. This reveals a statistically significant association between the geographic location of the respondents and their views on the primary beneficiary of generative AI—individuals or organizations. These results highlight regional differences in the expectations of generative AI's impact on individuals and organizations.

This segment of the analysis seeks to understand regional expectations regarding the timeline for the development of Artificial General Intelligence (AGI) systems. Participants were asked, "When do you estimate AGI systems will be built?" The results vary by region, with Asia and South America having a relatively larger percentage of respondents expecting AGI systems to be built within the next 2 to 5 years (31.0% and 37.5% respectively). In contrast, the greatest anticipation for AGI development within the next 5 to 10 years is in Australia, at 47.1%. There is also a notable skepticism about AGI ever being developed, with 13.5% of European respondents and 12.1% of North American respondents expressing this view.

| | | Where do you live? | | | | | |
|---|---|---|---|---|---|---|---|
| | | Europe | North America | Asia | South America | Africa | Australia |
| When do you estimate AGI systems will be built? | In the next 2 years | 8.4% | 6.0% | 8.1% | 9.4% | 8.0% | 5.9% |
| | In the next 2 to 5 years | 21.1% | 14.7% | 31.0% | 37.5% | 28.0% | 29.4% |
| | In the next 5 to 10 years | 25.5% | 29.3% | 27.1% | 21.9% | 36.0% | 47.1% |
| | In the next 10 to 20 years | 19.8% | 19.0% | 14.8% | 12.5% | 16.0% | 11.8% |
| | More than 20 years | 10.8% | 18.1% | 12.9% | 9.4% | 4.0% | 0.0% |
| | Never | 13.5% | 12.1% | 6.2% | 9.4% | 8.0% | 5.9% |

Statistical analysis via the chi-square test resulted in a Chi-square statistic of 38.509 with a p-value of approximately 0.0412 and 25 degrees of freedom. The p-value is below the standard alpha level of 0.05, allowing us to reject the null hypothesis. This indicates a statistically significant variation in expectations regarding the construction of AGI systems based on the geographic location of respondents. It suggests that regional factors may influence how soon people believe AGI will be a reality.

This analysis examines the relationship between respondents' industries and their geographic location. This relationship between demographic variables is important to understand the composition of respondents. Survey participants were asked about the sector in which they work, with options spanning from Communication Services to Utilities, including an "Other Sector" category. The data indicates distinct regional patterns in industry employment. Notably, the Financial sector has a strong presence in Africa (44.0%) and Australia (41.2%), while the Information Technology sector is significantly represented in North America (28.4%) and Asia (31.0%). The "Other Sector" category sees its highest proportion in South America at 40.6%.

| | | **Where do you live?** | | | | | |
|---|---|---|---|---|---|---|---|
| | | Europe | North America | Asia | South America | Africa | Australia |
| **What industry do you work in?** | Communication Services | 5.6% | 5.2% | 5.2% | 3.1% | 4.0% | 0.0% |
| | Consumer Discretionary | 5.4% | 8.6% | 3.3% | 6.3% | 0.0% | 0.0% |
| | Consumer Staples | 1.4% | 2.6% | 3.3% | 0.0% | 0.0% | 5.9% |
| | Energy | 2.9% | 0.9% | 4.3% | 3.1% | 12.0% | 0.0% |
| | Financials | 22.6% | 12.9% | 14.8% | 21.9% | 44.0% | 41.2% |
| | Health Care | 7.9% | 7.8% | 6.2% | 3.1% | 0.0% | 0.0% |
| | Industrials | 5.7% | 4.3% | 5.2% | 3.1% | 0.0% | 5.9% |
| | Information Technology | 18.6% | 28.4% | 31.0% | 9.4% | 16.0% | 17.6% |
| | Materials | 1.5% | 1.7% | 2.4% | 3.1% | 0.0% | 0.0% |
| | Real Estate | 1.1% | 0.9% | 1.4% | 3.1% | 8.0% | 0.0% |
| | Utilities | 0.5% | 0.0% | 0.5% | 3.1% | 0.0% | 5.9% |
| | Other Sector | 26.5% | 26.7% | 22.4% | 40.6% | 16.0% | 23.5% |

Statistical testing using the chi-square method yields a Chi-square statistic of 93.408 with a p-value of approximately 0.000947 and 55 degrees of freedom. With the p-value being substantially below the alpha level of 0.05, the null hypothesis can be rejected. This result indicates that there is a statistically significant association between the industry in which individuals work and their geographic location, pointing to regional specialization or variation in industrial representation across the globe.

**Discussion and Conclusion**

The research reported here aims to understand how new technologies like generative AI are quickly adopted and used by executive and managerial leaders to create value in organizations. A survey of INSEAD's global alumni base in the early diffusion of LLM technologies revealed several intriguing insights into perceptions and engagements with generative AI across a broad spectrum of demographics, industries, and geographies. Notably, there's a prevailing optimism about the role of generative AI in enhancing productivity and innovation, particularly as regards time savings and efficiency. Despite this enthusiasm and widespread adoption, concerns are significant, particularly regarding misuse by individuals and issues related to surveillance and privacy.

A particular strength of this research is the heterogeneous population of business leaders in the INSEAD alumni population which we sampled. Analysis revealed different attitudes across demographic variables. Younger respondents are significantly more excited about generative AI and more likely to be using it at work and in personal life than older participants. Those in Europe have a somewhat more distant view of generative AI than those in North America in Asia, in that they see the gains more likely to captured by organizations than individuals, and are less likely to be using it professional and personal contexts than those in North America and Asia.  This may also be related to the fact that those in Europe are more likely to be working in Financial Services and less likely to be working in Information Technology industries than those in North America and Asia. Despite this, those in Europe are more likely to see AGI happening faster than those in North America, although this may reflect less interaction with generative AI in personal and professional contexts. These findings collectively underscore the complex and multifaceted perceptions of generative AI's role in society, pointing to both its promising potential and the challenges it presents.

**Bibliography**

Agrawal, Ajay, Joshua S. Gans, and Avi Goldfarb. 2023. 'Do we want less automation?', *Science*, 381: 155-58.

Ali, Sanna J., Angèle Christin, Andrew Smart, and Riitta Katila. "Walking the Walk of AI Ethics: Organizational Challenges and the Individualization of Risk among Ethics Entrepreneurs." In *2023 ACM Conference on Fairness, Accountability, and Transparency (FAccT '23)*. Chicago, IL.

Beane, Matt. 2019. 'Learning to Work with Intelligent Machines.', *Harvard Business Review*, 97: 140-49.

Berg, J. M., M. Raj, and R. Seamans. 2023. 'Capturing value from artificial intelligence', *Academy of Management Discoveries*, 9: 424-28.

Bernd Carsten Stahl, Josephina Antoniou, Nitika Bhalla, Laurence Brooks, Philip Jansen, Blerta Lindqvist, Alexey Kirichenko, Samuel Marchal, Rowena Rodrigues, Nicole Santiago, Zuzanna Warso, David Wright 2023. 'A systematic review of artificial intelligence impact assessments', *Artificial Intelligence Review*, 56: 12799-831.

Boussioux, L., J. L. Jane, M. Zhang, V. Jacimovic, and K. Lakhani. 2023. "The crowdless future? How generative AI is shaping the future of human." In *Harvard Business School Technology & Operations Mgt. Unit Working Paper*.

Bresnahan, Timothy F., and Manuel Tratjenberg. 1995. 'General Purpose Technologies: 'Engines of Growth'?', *Journal of Econometrics*, 65: 83-108.

Brynjolfsson, Erik, Danielle Li, and Lindsey R. Raymond. 2023. "Generative AI at work." In *NBER Working Paper*.

Burt, Ronald. 2001. 'Attachment, decay, and social network', *Journal of Organizational Behavior*, 22: 619-43.

Chomsky, N., I. Roberts, and J. Watumull. 2023. 'Noam Chomsky: The False Promise of ChatGPT'.

Davis, Jason P. 2023. *Digital Relationships: Network Agency Theory and Big Tech* (Stanford Univeristy Press: Stanford, CA).

Dell'Acqua, F., E. McFowland, E. R. Mollick, H. Lifshitz-Assaf, K. Kellogg, and K. R. Lakhani. 2023. "Navigating the jagged technological frontier: field experimental evidence of the effects of AI on knowledge worker productivity and quality." In *Harvard Business School Technology & Operations Mgt. Unit Working Paper*.

Doshi, A. R., and O. P. Hauser. 2024. "Generative artificial intelligence enhances individual creativity but reduces the collective diversity of novel content." In.

Doshi, Anil Rajnikant and Bell, J. Jason and Mirzayev, Emil and Vanneste, Bart. 2024. 'Generative Artificial Intelligence and Evaluating Strategic Decisions', *Available at SSRN: https://ssrn.com/abstract=4714776 or http://dx.doi.org/10.2139/ssrn.4714776*.

Eesley, Charles E. 2011. 'Alumni surveys as a data collection methodology', *Working Paper*.

Eesley, Charles E, J.B. Li, and D. Yang. 2012. 'Does Institutional Change in Unversities Influence High-Tech Entrepreneurship? Evidence from China's Project 985', *Working Paper*.

Eloundou, Tyna, Sam Manning, Pamela Mishkin, and Daniel Rock. 2023. "Gpts are gpts: An early look at the labor market impact potential of large language models." In.

Felten, Edward, M. Raj, and R. Seamans. 2023. 'How will Language Models like ChatGPT Affect Occupations and Industries?'.

Gaessler, F., and H. Piezunka. 2023. 'Training with AI: evidence from chess computers', *Strategic Management Journal*, 44: 2724-50.

Girotra, Karan, Lennart Meincke, Christian Terwiesch, and Karl T. Ulrich. 2023. "Ideas are dimes a dozen: Large language models for idea generation in innovation." In.

Goldfarb, B. 2005. 'Diffusion of general-purpose technologies: understanding patterns in the electrification of US Manufacturing 1880-1930.', *Industrial and Corporate Change*, 14: 745-73.

Hannigan, Timothy, Ian P. McCarthy, and Andre Spicer. 2023. "Beware of Botshit: How to Manage the Epistemic Risks of Generative Chatbots." In.

Hui, Xiang and Reshef, Oren and Zhou, Luofeng. 2023. 'The Short-Term Effects of Generative Artificial Intelligence on Employment: Evidence from an Online Labor Market', *Available at SSRN: https://ssrn.com/abstract=4527336 or http://dx.doi.org/10.2139/ssrn.4527336*.

INSEAD. 2023. 'Maag INSEAD Centre for Entrepreneurship'.

———. 2024. https://www.insead.edu/alumni.

Jia, N., X. Luo, Z. Fang, and C. Liao. 2023. 'When and how artificial intelligence augments employee creativity', *Academy of Management Journal*.

KPMG. 2023. 'KPMG U.S. survey: Executives expect generative AI to have enormous impact on business, but unprepared for immediate adoption'. https://kpmg.com/us/en/media/news/kpmg-generative-ai-2023.html#:~:text=A%20transformative%20technology%20and%20competitive,than%20any%20other%20emerging%20technology.

Lazear, E. 2004. 'Balanced skills and entrepreneurship.', *American Economic Review*, 94: 208-11.

Lebovitz, S., N. Levina, and H. Lifshitz-Assaf. 2021. 'Is AI ground truth really true? The dangers of training and evaluating ai tools based on experts' know-what', *MIS Quarterly*, 45: 1501-25.

Lebovitz, Sarah, Hila Lifshitz-Assaf, and Natalia Levina. 2022. 'To engage or not to engage with AI for critical judgments: How professionals deal with opacity when using AI for medical diagnosis', *Organization Science*, 33: 126-48.

Lee, Peter, Carey Goldberg, and Isaac Kohane. 2023. *The AI Revolution in Medicine: GPT-4 and Beyond* (Pearson).

Mollick, Ethan R., and Lilach Mollick. 2023. "Assigning AI: Seven Approaches for Students, with Prompts." In.

Mukherjee, A., and H. H. Chang. 2023. 'Managing the Creative Frontier of Generative AI: The Novelty-Usefulness Tradeoff', *California Management Review Insights*.

Naumovska, Ivana, Vibha Gaba, and Henrich Greve. 2021. 'The Diffusion of Differences: A Review and Reorientation of 20 Years of Diffusion Research', *Academy of Management Annals*, 15: 377-405.

OpenAi. 2023. "GPT-4 Technical Report." In.: OpenAI.

Otis, N. G., R. Clarke, S. Delecourt, D. Holtz, and R. Koning. 2023. "The uneven impact of generative AI on entrepreneurial performance." In.

Raisch, S., and S. Krakowski. 2021. 'Artificial intelligence and management: The automation–augmentation paradox', *Academy of Management Review*, 46: 192-210.

Tripsas, Mary. 2009. 'Technology, Identity, and Inertia Through the Lens of "The Digital Photography Company', *Organization Science*, 20: 441-60.

Tushman, Michael L., and Philip Anderson. 1986. 'Technological Discontinuities and Organizational Environments', *Administrative Science Quarterly*, 31: 439-65.

Vanneste, B. S., and P. Puranam. 2024. 'Artificial Intelligence, trust, and perceptions of agency', *Academy of Management Review*.